%% file: ms.tex
\documentclass[twocolumn]{article}
\usepackage{sectsty}
\sectionfont{\fontsize{12}{15}\selectfont}
\usepackage[vlined]{algorithm2e}
\usepackage{color, enumitem}
\usepackage{amsmath}
\usepackage{amsthm}
\usepackage[vlined]{algorithm2e}
\usepackage{xcolor}
\usepackage{caption}
\usepackage{graphicx}
\usepackage{mathtools}
\usepackage{verbatim}
\usepackage{authblk}
\usepackage{blindtext}
\usepackage{hyperref}

\input{macros}

\title{A Differentially Private Wilcoxon Signed-Rank Test}

\author{Simon Couch}
\author{Zeki Kazan}
\author{Kaiyan Shi}
\author{Andrew Bray}
\author{Adam Groce}
\affil{\textit{\{couchs, kazanz, kaishi, abray, agroce\}@reed.edu}}
\affil{Reed College Mathematics Department}

\date{}

\begin{document}

\maketitle

\section{Abstract}

Hypothesis tests are a crucial statistical tool for data mining and are the workhorse of scientific research in many fields.  Here we present a differentially private analogue of the classic Wilcoxon signed-rank hypothesis test, which is used when comparing sets of paired (e.g., before-and-after) data values. We present not only a private estimate of the test statistic, but a method to accurately compute a p-value and assess statistical significance.  We evaluate our test on both simulated and real data.  Compared to the only existing private test for this situation, that of Task and Clifton, we find that our test requires less than half as much data to achieve the same statistical power.

\input{introduction.tex}

\input{Background.tex}

\input{Our_Algorithm.tex}

\input{Results.tex}

\input{Comparison.tex}

\input{Application.tex}

\input{Conclusion.tex}

\clearpage
\bibliography{sources}

\clearpage
\appendix
\input{Appendix.tex}

\end{document}

%% file: macros.tex
\let\svthefootnote\thefootnote
\newcommand\blankfootnote[1]{%
  \let\thefootnote\relax\footnotetext{#1}%
  \let\thefootnote\svthefootnote%
}

\newcounter{ag}
\addtocounter{ag}{1}

\newcounter{ab}
\addtocounter{ab}{1}

\newcounter{sc}
\addtocounter{sc}{1}

\newcounter{ks}
\addtocounter{ks}{1}

\newcounter{zk}
\addtocounter{zk}{1}

\newcommand{\data}{\ensuremath{\mathbf{x}} \xspace}
\newcommand{\Data}{\ensuremath{\mathbf{X}} \xspace}
\newcommand{\lap}{\ensuremath{\textsf{Lap}} \xspace}

\newcommand{\privW}{\ensuremath{\mathcal{\widetilde{W}}} \xspace}
\newcommand{\WP}{\ensuremath{\mathcal{WP}} \xspace}
\newcommand{\privWP}{\ensuremath{\mathcal{\widetilde{WP}}} \xspace}
\newcommand{\complete}{\ensuremath{\mathcal{\widetilde{C}}} \xspace}

\newcommand{\algrule}[1][.8pt]{\par\vskip.1\baselineskip\hrule height #1\par\vskip.1\baselineskip}

\newtheorem{theorem}{Theorem}[section]

\theoremstyle{definition}
\newtheorem{definition}[theorem]{Definition}
\newtheorem{example}[theorem]{Example}

\newcommand{\rom}[1]{\uppercase\expandafter{\romannumeral #1\relax}}

%% file: introduction.tex
\section{Introduction}

Consider a scenario where a medical researcher measures a patient's blood pressure before and after a particular intervention.  Once the data is obtained, the researcher wants to determine whether the intervention had any effect, or whether the difference between the two samples was just random variation.  The data generated in this study is an example of \emph{paired-sample} data, where there exists a natural link between the two measurements of the same individual. This is a common form of data in scientific research, and one of the traditional methods to check for the signficance of an effect is the Wilcoxon signed-rank test \cite{WilcoxonDefinition}.  This test does not require any additional assumptions about the data (e.g., that it was normally distributed), and so it can be used on social network data and other data for which it may not be appropriate to assume normality.

Unfortunately, in many of the use cases for the Wilcoxon test, the data in question is private and sensitive (e.g., medical records, financial records or social network data).  Ethical and legal concerns prevent the easy sharing of this data.  Frequently, research cannot be done because the owners of existing data are not willing to accept the risk inherent in entrusting this data to a third party.  Even if the researcher is entrusted with access to the data, the researcher themselves must now decide what they can publish.  The release of even just summary statistics has led to privacy violations \cite{homer2008resolving}, but without the release of some results the research is pointless.

Differential privacy provides a criterion that is sufficient to provably guarantee that the results of a particular query on a database do not violate individual privacy.  Differential privacy generally requires the query be randomized, with the output usually aiming to approximate the result of a non-private query as accurately as possible.  Years of research have resulted in a variety of effective algorithms for differentially private queries, allowing the approximation of everything from histograms \cite{DPDefinitions} to regression model coefficients \cite{Sheffet2015,LogisticRegression,LinearRegression} to classifiers from machine learning algorithms \cite{Classifiers2,Classifiers}.

In this paper we provide an algorithm that carries out a private analogue of the Wilcoxon signed-rank test.  A private version of the test will always have lower statistical power than the classical version, meaning that more data is needed before a given effect can be distinguished from random noise.  The first differentially private analogue of the Wilcoxon test was given by Task and Clifton in 2016 \cite{DPWilcoxon}.  Here we strive to improve on that result, introducing a new test with higher power.

In a hypothesis test, a test statistic is computed and compared to a reference distribution to see if it is abnormally high.  The difficulty of developing a private test comes not just from the need to privately approximate a test statistic, but also from the need to rigorously define ``abnormally high.''  It is not sufficient to treat the approximate test statistic as equivalent to its non-private counterpart.

The key contributions of this work are as follows:

\paragraph{A new private test statistic}  We estimate not the standard Wilcoxon test statistic but a variant introduced by Pratt in 1959 \cite{Pratt}.  We argue that this variant is more amenable to being privatized without losing its utility.  We prove a bound on its sensitivity and provide a private algorithm for approximating this statistic.

\paragraph{Accurate p-value calculation}  We show how to calculate an accurate p-value that takes into account the noise added for privacy.  We do this through simulation, numerically approximating (to a high degree of accuracy) the relevant reference distribution.  This is a more exact method than the upper bounds given in prior work.

\paragraph{Experimental implementation}  We implement our algorithm and measure its power.  We also implement the test of Task and Clifton and along the way correct a significant error in their work.  We then compare the two tests and find that ours can detect the same effect with only 40\% of the data needed previously, closing the majority of the gap between the private and public versions of the test.  The source code for our implementation is publicly available.\footnote{Our code is available at: \href{http://www.github.com/simonpcouch/wilcoxon}{http://www.github.com/simonpcouch/wilcoxon}}

%% file: Background.tex
\section{Background} \label{background}

We begin by outlining some background on hypothesis testing, followed by some specifics on the Wilcoxon signed-rank test. We then discuss differential privacy and previous applications of differentially private hypothesis testing.

\subsection{Hypothesis Testing}

Hypothesis testing is a very common form of data analysis, particularly in applied scientific research.  The goal of a hypothesis test is to measure whether a particular data set is consistent with a given theory describing how the data is generated.  This theory is called the \textit{null hypothesis} ($H_0$).  For example, given blood pressure data before and after some treatment, researchers might want to determine if the change they see could plausibly be consistent with a null hypothesis that the treatment has no effect and the observed effect is due only to sampling variability.

%The first step in making this determination is to calculate a \textit{test statistic}, which is a function of the database, $T=f(\Data)$. The database, $\Data$, is conceived of as being drawn randomly according to $H_0$, and therefore $T$ itself is follows a distribution that can be determined analytically, through simulation, or through large sample approximation.

To make this determination, a \textit{test statistic}, $t$, is computed.  
%A particular hypothesis test is largely determined by the function used to compute the test statistic.  
Given a function $f$ for computing a test statistic, one can determine analytically, through simulation, or via large sample approximation, the distribution that the statistic $T=f(\Data)$ would have when the database $\Data$ is drawn randomly according to $H_0$.\footnote{We use the convention of denoting random variables with a capital roman letter and a particular value taken by that random variable with a lowercase roman letter.}  One can then compute a \emph{p-value}, defined as the probability that a test statistic as or more extreme than that observed would occur under $H_0$.

\begin{definition}
For an observed database \data, the test statistic $t=f(\data)$, and a null hypothesis $H_0$, the \emph{p-value} is defined as
$$\Pr[T \geq t \mid T = f(\Data) \text{ and } \Data \leftarrow H_0].$$
\end{definition}

Low p-values indicate that the test statistic is more extreme than is likely by random chance and may cause a researcher to reject $H_0$ as a plausible explanation of the data.  Normally a threshold value $\alpha$ is chosen and $H_0$ is rejected if $p<\alpha$. The chosen value for $\alpha$ determines the \textit{type \rom{1} error rate}, the probability of incorrectly rejecting if the null hypothesis is true. The value of $t$ at which $p=\alpha$, which we will denote $t^*$, is called the \textit{critical value}.

Hypothesis tests are judged by their \textit{statistical power}.  This is a measure of how likely the test is to detect an effect and reject the null hypothesis when it is indeed false.  Statistical power is a function of the amount of data and the particular alternate distribution (i.e., the effect size).

\begin{definition}
For a given alternate data distribution $H_A$, the \emph{statistical power} of a hypothesis test is 
$$\Pr[T \geq t^* \mid T = f(\Data) \text{ and } \Data \leftarrow H_A].$$
\end{definition}

A primary goal of hypothesis test design is to find tests with high power.  In classical statistics there are many hypothesis tests that have been proven to be optimal for a large set of use cases.

%FROM ADAM: Several prior versions were mixed up in comments below.  I removed them because the different versions were indistinguishable.  Use the repo to find old versions.

\subsection{Wilcoxon Signed-Rank Test}\label{wc}

Consider again the example where researchers want to see if a treatment caused a change in blood pressure. A standard hypothesis test for this situation is the Wilcoxon signed-rank test,  proposed in 1945 by Frank Wilcoxon \cite{WilcoxonDefinition}. This test evaluates the difference between values obtained from two paired samples (such as subjects before and after treatment, husbands and wives, etc.) to see if they plausibly come from the same distribution. The test assumes that all pairs are independent, random draws from a distribution with an ordinal scale. In the public setting, when the data is known to be normally distributed, the t-test consistently outperforms the Wilcoxon. However, the Wilcoxon test does not assume normality, which is beneficial when the underlying distribution of the sample data is not known.

The function calculating the Wilcoxon test statistic is formalized in Algorithm $\mathcal{W}$.  Given a database $\data$ containing sets of pairs $(u_i, v_i)$, the test computes the difference $d_i$ of each pair, drops any with $d_i=0$, and then ranks them by magnitude.  (If magnitudes are equal for several differences, all are given a rank equal to the average rank for that set.)   If $s_i = \pm 1$ is the sign of $d_i$ and $r_i$ is its rank, then  $w = \sum_i s_i r_i$.

\begin{algorithm}
\DontPrintSemicolon
\algrule
\textbf{Algorithm } $\mathcal{W}$ \textbf{:} Wilcoxon Test Statistic Calculation \;
\algrule
\KwIn{$\data$}
\Begin{
    \For{row $i$ of $\data$}{
        $d_i \longleftarrow |v_i - u_i|$\;
        $s_i \longleftarrow \textup{Sign}(v_i - u_i)$\;
    }
    Order the terms from lowest to highest $d_i$\;
    Drop any $d_i = 0$\;
    \For{row $i$ of $\data$}{
        $r_i \longleftarrow \textup{rank of row }i$
   }
    %\For{$d_a$ = $d_b = \ldots = d_n$ \textup{for some} $a, \ldots , n$}{
       % $R_a,\ldots R_n \longleftarrow \textup{the average row rank of } a, \ldots, n$\;
      %  }
   $w \longleftarrow \sum_i s_i r_i$
   
}
\KwOut{$w$}
\algrule
\label{alg:wc}
\end{algorithm}

\begin{example}
Table \ref{tab:i} presents an example database containing five data elements.  The values $u_i$ and $v_i$ are inputs in the database and $d_i$ and $s_i$ are the results of the initial stage of computation.

\begin{table}[h!]
\centering
\caption{Initial Table} \label{tab:i}
\begin{tabular}{ccccc}
$i$ & $u_i$ & $v_i$ & $d_i$ & $s_i$ \\
\hline
1   & 9     & 18    & 9 & 1     \\
2   & 2     & 11    & 9 & 1   \\
3   & 3     & 3     & 0 & -   \\
4   & 8     & 10    & 2 & 1     \\
5   & 9     & 8     & 1 & -1  
\end{tabular}
\end{table}

In table \ref{tab:f} we sort the elements in increasing order by $d_i$.  We drop $i = 3$ because $d_3=0$.  We then compute ranks.  Because there is a tie for ranks 3 and 4, we set the rank of these values to 3.5.

\begin{table}[h!]
\centering
\caption{Final Table} \label{tab:f}
\begin{tabular}{cccc}
$i$ & $d$ & $s$ & $r$ \\
\hline
5    & 1     & -1   & 1     \\
4    & 2     & 1    & 2     \\
1    & 9     & 1    & 3.5   \\
2    & 9     & 1    & 3.5  
\end{tabular}
\end{table}

We then compute the test statistic $$w = \sum_i s_i r_i = 3.5 \cdot 1 + 3.5\cdot 1 + 2 \cdot 1 + 1 \cdot (-1) = 8.$$ 
\end{example}

Under the null hypothesis that $u_i$ and $v_i$ are drawn from the same distribution, the distribution of the test statistic $W$ can be calculated exactly using combinatorial techniques. This becomes computationally infeasible for large databases, but an approximation exists in the form of the normal distribution with mean 0 and variance $\frac{n_r(n_r+1)(2n_r+1)}{6}$, where $n_r$ is the number of rows that were not dropped.  Knowing this, one can calculate the p-value for any particular value of $w$.

\subsection{Differential Privacy}

When working with data that contains sensitive information about individuals, the need to publicly share the result of the hypothesis test must be balanced with the need to protect privacy. Unfortunately, traditional methods of handling sensitive data provide poor privacy protections.  Databases privatized with ad hoc anonymization techniques (replacing names with unique numeric identifiers, for example) have frequently been attacked, resulting in substantial privacy violations \cite{AOL,Sweeney2002,NYCData}. Even summary statistics can allow malicious actors to violate the privacy of people in the database \cite{homer2008resolving}.

Differential privacy \cite{DPDefinitions} is a definition that, when satisfied, provably guarantees that the output of a query cannot be used to infer anything about the individuals who contributed to the database.  Informally, it requires that for any individual, the query output would be roughly the same for any value of that individual's data.  To do this, the query mechanism must be randomized.  More formally:

\begin{definition}[Differential Privacy] \label{def:dp}
 A randomized algorithm $\tilde{f}$ on databases is $\epsilon$-differentially private if for all $\mathcal{S} \subseteq \textup{Range}(\tilde{f})$ and
for databases $\data, \data\mathbf{'}$ that only differ in one row
$$\Pr[\tilde{f}(\data) \in \mathcal{S}] \leq \textup{exp}(\epsilon) \Pr[\tilde{f}(\data\mathbf{'}) \in \mathcal{S}].$$
\end{definition}

We say that $\data$ and $\data\mathbf{'}$ that differ only in one row are \textit{neighboring databases} and we call $\epsilon$ the \textit{privacy parameter}. The user selects $\epsilon$, which determines the strength of the privacy. 

One useful property of differential privacy is resistance to post-processing, which ensures that no function of a differentially private algorithm can cause a privacy loss \cite{DPDefinitions}.

\begin{theorem}[Post Processing] \label{thm:pp}
 Let $\tilde{f}$ be an $\epsilon$-differentially private randomized
algorithm. Let $g$ be an
arbitrary randomized algorithm. Then $g \circ \tilde{f}$ is $\epsilon$-
differentially private.
\end{theorem}

Resistance to post-processing is an important theorem for two reasons.  First, any good privacy definition must have this property (or something very similar).  If it didn't, then the privacy of the output is only superficial and some sort of analysis can extract the private information.  Secondly, this shows that post-processing can be used to design private algorithms.  If one step of a computation is differentially private, then as long as the rest of the computation does not directly access the database, the result of the full computation will be private as well.  We use this technique later.

One standard technique for creating differentially private algorithms is the \textit{Laplace mechanism}, introduced by Dwork et al. in 2006 \cite{DPDefinitions}.  Given an arbitrary (non-private) function $f$, the Laplace mechanism provides one way to create a private function $\tilde{f}$ that approximates $f$.  To do this, one must first compute the sensitivity of $f$, meaning the maximum effect a change in a single row can have on the output.

\begin{definition}[Sensitivity] \label{def:gs}
 The sensitivity of a function $f$  is
$$\Delta f = \underset{\data,\data\mathbf{'}}{ \textup{max}}\ |f(\data)-f(\data\mathbf{'})|,$$ where $\data$ and $\data\mathbf{'}$ are neighbouring databases. 
\end{definition}

The Laplace mechanism adds noise to $f$, where the noise is sampled from a Laplace distribution, a double-sided exponential distribution.

\begin{definition}[Laplace Distribution]\label{def:ld}
 The Laplace distribution
(centered at 0) with scale $b$ is the distribution with probability density function:$$ \lap(x|b)=\frac{1}{2b}\textup{exp}\Big(-\frac{|x|}{b}\Big).$$ We usually write $\lap(b)$ to denote the Laplace distribution with scale $b$.
\end{definition} 

The Laplace mechanism simply adds noise produced from the Laplace distribution to the output of a function. The magnitude of the noise is determined by the choice of $\epsilon$ and the sensitivity of the function.

\begin{definition}[Laplace Mechanism]  \label{def:lm}
Given any function $f$, the Laplace mechanism is defined as:$$\tilde{f}(\data)= f(\data)+Y,$$ where $Y$ is drawn from $\lap(\Delta f/\epsilon)$, and $\Delta f$ is the sensitivity of $f$.
\end{definition}

The Laplace mechanism is one of the most commonly used methods to achieve differential privacy.  Dwork et al.~\cite{DPDefinitions} proved the following:

\begin{theorem} \label{thm:lm}
(Laplace Mechanism) The Laplace mechanism preserves $\epsilon$-differential privacy.
\end{theorem}

\input{DPHT.tex}

\input{related_dpht.tex}

%% file: DPHT.tex
\subsection{Differentially Private Hypothesis Tests}

To carry out a differentially private hypothesis test, one must first find a differentially private function $\tilde{f}$ that computes a useful test statistic.  This could be either a differentially private estimate of a standard test statistic, or a completely new test statistic.\footnote{We note here that our test statistics are now randomized, rather than deterministic functions of the database.  This actually causes very few mathematical issues, but it does mean that the output of a hypothesis test on fixed data will not be the same on each run.}

In a testing framework, a test statistic is \textit{not} a meaningful output on its own; it is an intermediate calculation on the way to a p-value. One cannot simply use the estimated statistic in place of the true one and carry out a p-value computation as in the non-private setting, as resultant p-values can be wildly innacurate and artificially inflate the type I error rate \cite{ANOVA,Wang2015}.  Instead, one must compute a new reference distribution modeling the distribution of $\tilde{f}(\Data)$ when $\Data$ is generated under $H_0$.

A fully developed differentially private hypothesis test includes not just a method for computing the test statistic, but also for computing the associated p-value.  The goal is to develop such a hypothesis test and to achieve power as close as possible to what can be achieved in the classic non-private setting.  Just as each type of input data requires a separate hypothesis test in the classical case, each type of input data requires its own differentially private test.

%% file: related_dpht.tex
\subsection{Related Work}

There is a significant body of work on differentially private hypothesis testing, broadly construed, but much of it takes a different approach than this work.  For example, some results (e.g., \cite{smith2008efficient, wasserman2010statistical, smith2011privacy}) look at how quickly various private approximations of test statistics converge to their limiting distributions.  These are important theoretical results, but they do not bring us to the point of practical, implementable tests.  There is often no discussion of a reasonable reference distribution when the noise is not yet negligible, and the results are often purely asymptotic, hiding potentially problematic constants.
%note: smith2008 is an earlier, unpublished version of smith2011

The hypothesis test for which private variants have been the most well-studied is the chi-squared test, which tests whether two categorical variables could plausibly be independent of each other.\footnote{There are variants of this test that can be used in slightly different situations, like seeing of a certain set of categorical data is consistent with a specified distribution.}    Vu and Slavkovi\'{c} \cite{vu2009differential} give differentially private versions of a single proportion test and a chi-squared test with clinical trial data in mind.  They give accurate p-value calculations that adjust for the added Laplace noise. Several studies of private chi-squared tests have been specifically intended for use with genome-wide association study (GWAS) data \cite{fienberg2011privacy, uhlerop2013privacy, johnson2013privacy}.  These papers make asymptotic arguments for the reliability of their p-value calculations.  Other work has used Monte Carlo simulations to achieve higher precision \cite{gaboardi2016differentially, wang2015revisiting}.  Rogers and Kifer \cite{rogers2017new} instead proposes a new statistic with an asymptotic distribution more similar to its non-private analogue. While all of these papers discuss private test statistics and many (but not all) discuss the distribution of that statistic under the null hypothesis, few carefully demonstrate the statistical power of the resulting test.  Rogers and Kifer \cite{rogers2017new} and Gaboardi et al.~\cite{gaboardi2016differentially} are notable exceptions, giving power curves for several different approaches.

Methods for numerical (rather than categorical) data are less well-developed.  Some work has addressed testing the value of a mean or the difference of means \cite{solea2014differentially, d2015differential, ding2018comparing}.  While there is a large body of work on differentially private linear regression that aims to output an accurate best-fit line, only recently has work been done on using coefficients as test statistics \cite{sheffet2015differentially, barrientos2017differentially}.  (Such inference is near-ubiquitous in many academic fields.)  Campbell et al.~\cite{ANOVA} give a private analogue of a one-way analysis of variance (ANOVA) test, and Nguy{\^e}n and Hui propose a test for surival analysis data \cite{nguyen2017differentially}.

%% file: Our_Algorithm.tex
\section{Our Algorithm}

In this section, we will define our algorithm for the differentially private Wilcoxon test and prove that it is $\epsilon$-differentially private.  At a high level, our algorithm is quite straightforward; we compute a test statistic as one might in the public case and add Laplacian noise to make it private.  However, there are several important innovations that greatly increase the power of our test.

Our first innovation is to use a different variant of the Wilcoxon test statistic.  While the version introduced in Section \ref{wc} is the one most commonly used, other versions have long existed in the statistics literature.  In particular, we look at a variant introduced by Pratt in 1959 \cite{Pratt}.  In this variant, rather than dropping rows with $d_i = 0$, those rows are included.  When $d_i=0$ we set $s_i = \textup{Sign}(d_i) = 0$, so those rows contribute nothing to the resultant statistic, but they do push up the rank of other, non-tied rows.  We define the algorithm for this calculation, which we denote $\WP$, below.

\begin{algorithm}
\DontPrintSemicolon
\algrule
\textbf{Algorithm }$\mathcal{WP}$\textbf{:} Wilcoxon Statistic - Pratt Variant \;
\algrule
\KwIn{$\data$}
\Begin{
    \For{row $i$ of $\data$}{
        $d_i \longleftarrow |v_i - u_i|$\;
        $s_i \longleftarrow \textup{Sign}(v_i - u_i)$ \;
    }
    Order the terms from lowest to highest $d_i$\;
    \For{row $i$ of $\data$}{
        $r_i \longleftarrow \textup{rank of row }i$
   }
    %\For{$d_a$ = $d_b = \ldots = d_n$ \textup{for some} $a, \ldots , n$}{
       % $R_a,\ldots R_n \longleftarrow \textup{the average row rank of } a, \ldots, n$\;
      %  }
   $w \longleftarrow \sum_i s_i r_i$
   
}
\KwOut{$w$}
\algrule
\label{alg:wcp}
\end{algorithm}

This procedure is almost the same as that for the calculation of the standard Wilcoxon test statistic, but the small difference is crucial.  In the sample database given in \ref{tab:i}, the Pratt variant produces the values in Table \ref{tab:pratt}.

\begin{table}[h!]
\centering
\caption{Final Table (Pratt)} \label{tab:pratt}
\begin{tabular}{cccc}
$i$ & $d$ & $s$ & $r$ \\
\hline
3    & 0     & 0   & 1     \\
5    & 1     & -1   & 2     \\
4    & 2     & 1    & 3     \\
1    & 9     & 1    & 4.5   \\
2    & 9     & 1    & 4.5  
\end{tabular}
\end{table}
The resulting statistic is 
$$w = \sum_i s_i r_i = 4.5 \cdot 1 + 4.5\cdot 1 + 3 \cdot 1 + 2 \cdot (-1) = 10.$$

In the public setting, the Pratt variant is not very different from the standard Wilcoxon, being slightly more or less powerful depending on the exact effect one is trying to detect \cite{WCcomp}.  In the private setting, however, the difference is substantial.

The benefit to the Pratt variant comes from how the test statistics are interpreted.  In the standard Wilcoxon, it is known that the test statistic follows an approximately normal distribution, but the variance of that distribution is a function of $n_r$, the number of non-zero $d_i$ values.  In the private setting, this number is not known, and this has caused substantial difficulty in prior work.  (See Section \ref{comp_sec} for more discussion.)  On the other hand, the Pratt variant produces a test statistic that is always compared to the same normal distribution, which depends only on $n$.

\subsection{The Private Test Statistic}
Recall that $\data$ is a database and $n$ is the size of the database. The \WP algorithm (above) produces the public version of the Pratt-variant Wilcoxon statistic. The algorithm \privWP that outputs the differentially private analog is shown below. 

\begin{algorithm}[!htb]
\DontPrintSemicolon
\algrule
\textbf{Algorithm } $\privWP$ \textbf{:} Private Wilcoxon Statistic\;
\algrule
\KwIn{$\data$, $\epsilon$}
\Begin{
    $n \longleftarrow$ the number of paired samples in $\data$\;
    $w \longleftarrow \WP(\data)$\;
    $\widetilde{w} \longleftarrow w + \lap \Big (\frac{2n}{\epsilon} \Big )$\;
}
\KwOut{$\widetilde{w}$}
\algrule
\label{alg:wcalg}
\end{algorithm}

\input{Proof_of_wcalg_privacy.tex}

As we discussed previously, a test statistic on its own is not useful for statistical inference.  Given the test statistic, we must calculate a p-value. In the public case, we compare to a simple normal distribution, but in the private case the distribution of $\widetilde{w}$ under the null distribution is equal to the sum of a normal and a Laplacian.  This is difficult to compute analytically, so we simply simulate a large number $c$ of draws from this distribution. (In our experiments, $c$ is set to 10 million.  This gives a very accurate estimate of $p$.)  We combine everything into a final test algorithm in Algorithm \complete.

\begin{algorithm} [!htb]
\DontPrintSemicolon
\algrule
\textbf{Algorithm } $\complete$ \textbf{:} Complete Wilcoxon Test\;
\algrule
\KwIn{\data, $\epsilon$, $c$}
\Begin{
	$\widetilde{w} := \privWP(\data, \epsilon)$\;
    \For{$k := 1$ \textup{to} $c$}{
                $W_k \longleftarrow {\sf Normal}(0, n(n+1)(2n+1)/6) + \lap(2n/\epsilon)$;
    }
    $p \longleftarrow$ fraction of $W_k$ more extreme than $\widetilde{w}$ \;
}
\KwOut{$\widetilde{w}, p$}
\algrule
\label{alg:complete}
\end{algorithm}

In this algorithm we draw our reference distribution (the $W_k$ values) assuming there are no $d_i=0$ rows.  The distribution will technically differ slightly when there are many rows with $d_i=0$, but the difference is inconsequential in all but very extreme circumstances, and even then the calculations are overly conservative.  See Appendix \ref{uniformp} for more details.

\begin{theorem}
Algorithm \complete is $\epsilon$-differentially private. 
\end{theorem}

\begin{proof}
The computation of $\widetilde{w}$ was already shown to be private.  The remaining computation needed to find the p-value does not need access to the database---it is simply post-processing.  By Theorem \ref{thm:pp}, it follows that the complete algorithm is also private.
\end{proof}

%% file: Proof_of_wcalg_privacy.tex
\subsection{Proof of Differential Privacy}

\begin{theorem} \label{thm:wcalg}
Algorithm \privWP  is $\epsilon$-differentially private. 
\end{theorem}

\begin{proof}

It is sufficient to show that the sensitivity of \WP is bounded by $2n$.  The privacy of the algorithm then follows directly from Theorem \ref{thm:lm}.

\begin{figure} [!htb]
    \centering
    \includegraphics[width=\linewidth]{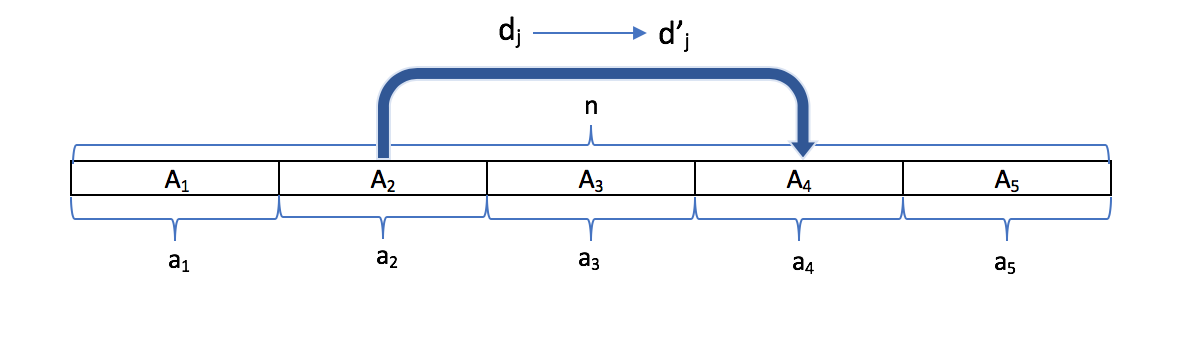}
    \caption{ $\data$ and $\data\mathbf{'}$ differ only in that $d_j$ is "moved" from region $A_2$ to region $A_4$.} \label{fig:Ggs}
\end{figure}

Let $\data$ and $\data\mathbf{'}$ be neighbouring databases. Let row $j$ be the row that differs between $\data$ and $\data\mathbf{'}$. Let $d_j$ and $d'_j$ indicate the difference between the data values in row $j$ in $\data$ and $\data\mathbf{'}$, respectively. Similarly, let $r_j$ and $r'_j$ indicate the rank of row $j$ in $\data$ and $\data\mathbf{'}$.  We assume without loss of generality that $d'_j > d_j$.

We divide the rows of $\data$ (or $\data\mathbf{'}$) into five regions, $A_1$ through $A_5$.  $A_1$ is the set of rows $i$ with $d_i < d_j$.  $A_2$ is the set with $d_i=d_j$.  $A_3$ has $d_i$ values between $d_j$ and $d'_j$.  $A_4$ has $d_i$ values equal to $d'_j$.  $A_5$ has values greater than $d'_j$ (though we don't need to use $A_5$ in the proof).  For convenience, we use $a_i$ to represent the size of the set $A_i$ in $\data$.

As shown in Figure \ref{fig:Ggs}, $\data$ and $\data\mathbf{'}$ differ only in that $d_j$ is ``moved'' from region $A_2$ to region $A_4$. Let $w$ and $w'$ be test statistics produced from databases $\data$ and $\data\mathbf{'}$, respectively.

Recall that $w = \sum_i s_i r_i$.  Let $w_i = s_i r_i$ be the contribution of row $i$ to this sum.  We then want to bound $\Delta w= |w-w'|$. We similarly set $\Delta w_i$ to be $|s_i r_i - s'_i r'_i|$.  Note that for $i$ in $A_1$ or $A_5$ we have $\Delta w_i = 0$.

Next, we consider $\Delta w_j$ the change in the term corresponding to the row where the data changes. As row $j$ is tied with all other rows in region $A_2$, $r_j = \frac{(a_1+1)+(a_1+a_2)}{2} = \frac{2a_1+a_2+1}{2}$ and $r'_j = \frac{(a_1+a_2+a_3)+(a_1+a_2+a_3+a_4)}{2} = \frac{2a_1+2a_2+2a_3+a_4}{2}$. In the worst case, $s_j$ changes when the move occurs, so 
\begin{align*}
\Delta w_j < r_j + r'_j &= \frac{2a_1+a_2+1}{2}+\frac{2a_1+2a_2+2a_3+a_4}{2} \\
&= \frac{4a_1+3a_2+2a_3+a_4+1}{2}.
\end{align*}

Now we consider $\Delta w_i$ where $i \in A_2$ (excluding row $j$). As there is one fewer row tied, each of their ranks (excluding row $j$) is changed from $\frac{2a_1+a_2+1}{2}$ to $\frac{2a_1+a_2}{2}$, which means that each row is changed by $\frac{1}{2}$. As there are $a_2-1$ rows in this region, the contribution to $\Delta w$ from this set of rows is bounded by $\frac{a_2-1}{2}$.

Similarly, each of the row ranks (excluding row $j$) in region $A_4$ is changed from $\frac{2a_1+2a_2+2a_3+a_4+1}{2}$ to $\frac{2a_1+2a_2+2a_3+a_4}{2}$ as one more row is tied with rows in $A_4$. As there are $a_4$ rows, $\Delta w$ from the rows in region $A_4$ is bounded by $\frac{a_4}{2}$. 

Now we consider region $A_3$. As $d_j$ has moved from before this region to after this region, each of the row ranks is changed by $1$. Therefore the contribution to $\Delta w$ from the rows in region $A_3$ is bounded by $a_3$.

It follows that 
\begin{align*}
    \Delta w &= |w-w'|\\
    &\leq \frac{4a_1+3a_2+2a_3+a_4+1}{2} + \frac{a_2-1}{2} + \frac{a_4}{2} + a_3\\
    &= 2a_1+2a_2+2a_3+a_4\\
    &\leq 2(a_1+a_2+a_3+a_4)\\
    &\leq 2n
\end{align*}

Having bounded the sensitivity, the proof is complete.

\end{proof}

%% file: Results.tex
\section{Experimental Results}\label{results}

We assess the power of our differentially-private Wilcoxon signed-rank test first on synthetic data.  (For tests with real data, see Section \ref{realdata}.) In order to measure power, we must first fix an effect size.  We chose to have the $u_i$ and $v_i$ values both generated according to normal distributions with means one standard deviation apart. We then measure the statistical power of Algorithm \complete (for a given choice of $n$ and $\epsilon$) by repeatedly randomly sampling a database \data from that distribution and then running \complete on that database.\footnote{Our actual implementation differs slightly from this.  To save time when running a huge number of tests with identical $n$ and $\epsilon$, we first generate the reference distribution $W_k$ values, which can be reused across runs.}  The power is the percentage of the time \complete returns a p-value less than $\alpha$.  (We use $\alpha=.05$ in all our experiments.)

\begin{figure} [!htb] 
    \centering
    \includegraphics[width=\linewidth]{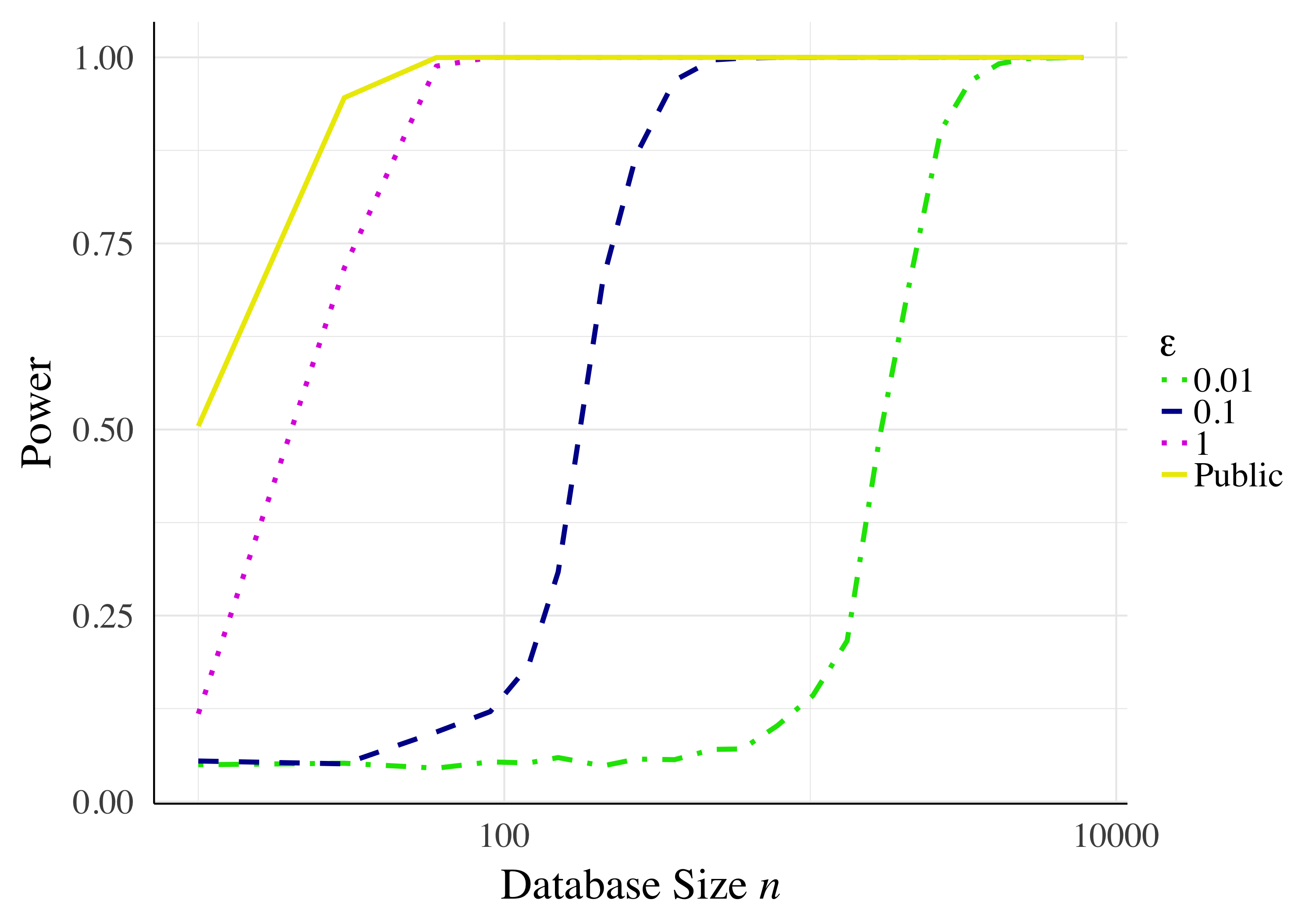}
    \caption{Power at various values of $\epsilon$ and sample size $n$. (Effect size: $\mu_u - \mu_v = 1 \sigma$; $\alpha = .05$)}
    \label{fig:Gepi}
\end{figure}

We consider several values of $\epsilon$.  The lowest, .01 is an extremely conservative privacy parameter and allows for safe composition with many other queries of comparable $\epsilon$ value. In Figure \ref{fig:Gepi}, we also use $\epsilon$s of $.1$ and $1$, which, while higher, still provide very meaningful privacy protection.  Ultimately, the choice of $\epsilon$ is a question of policy and depends on the relative weight placed on privacy and utility.  We also measure for comparison the power of the standard, public Wilcoxon test.\footnote{We use the standard version of the test because it is the one usually used.  The Pratt variant has very similar power.} As one might expect, the amount of data needed to detect a given effect scales roughly proportionately with $1/\epsilon$. 

\begin{figure} [!htb] 
    \centering
    \includegraphics[width=\linewidth]{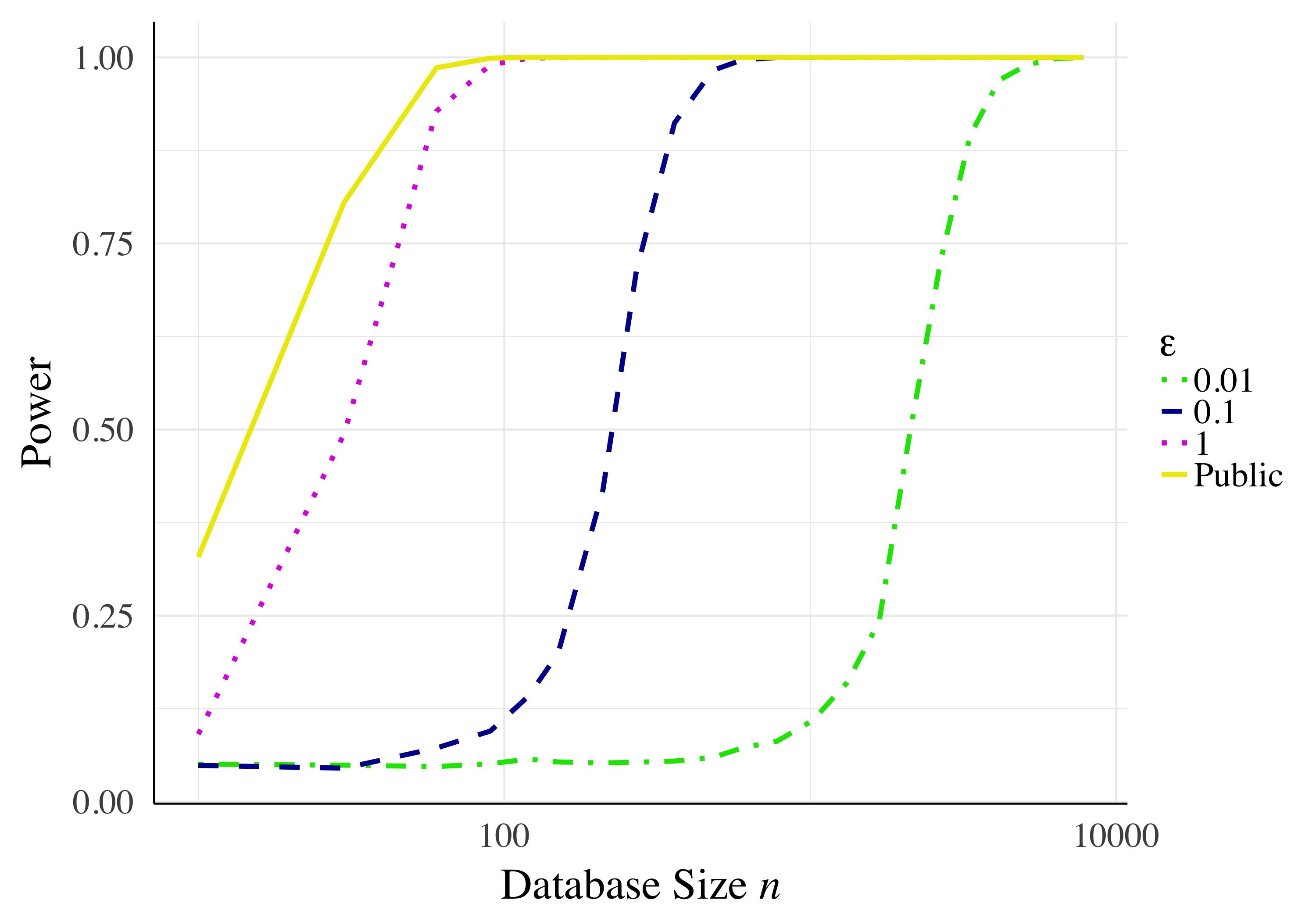}
    \caption{Power at various values of $\epsilon$ and sample size $n$. (Effect size: $\mu_u - \mu_v = 1 \sigma$; $\alpha = .05$; proportion of zeroes in $d_i$ is $.3$)}
    \label{fig:30_pct_ties}
\end{figure}

An important part of the algorithm design is the treatment of rows with $d_i=0$.  In Figure \ref{fig:30_pct_ties}, 30\% of rows have $d_i=0$ and the other 70\% are distributed as before, with the two data points sampled from normal distributions with means one standard deviation apart.  We find that while (as expected) all tests lose power, but they lose very little and the relative relationship between our tests and the non-private Wilcoxon test is unchanged.

In a properly calibrated hypothesis test, the distribution of the p-values under $H_0$ will be uniform, indicating that the Type I error rate is exactly $\alpha$. We verified that Algorithm \complete yields uniform p-values under $H_0$ in realistic settings. See Appendix \ref{uniformp} for further discussion.

\begin{figure} [!htb] 
    \centering
    \includegraphics[width=\linewidth]{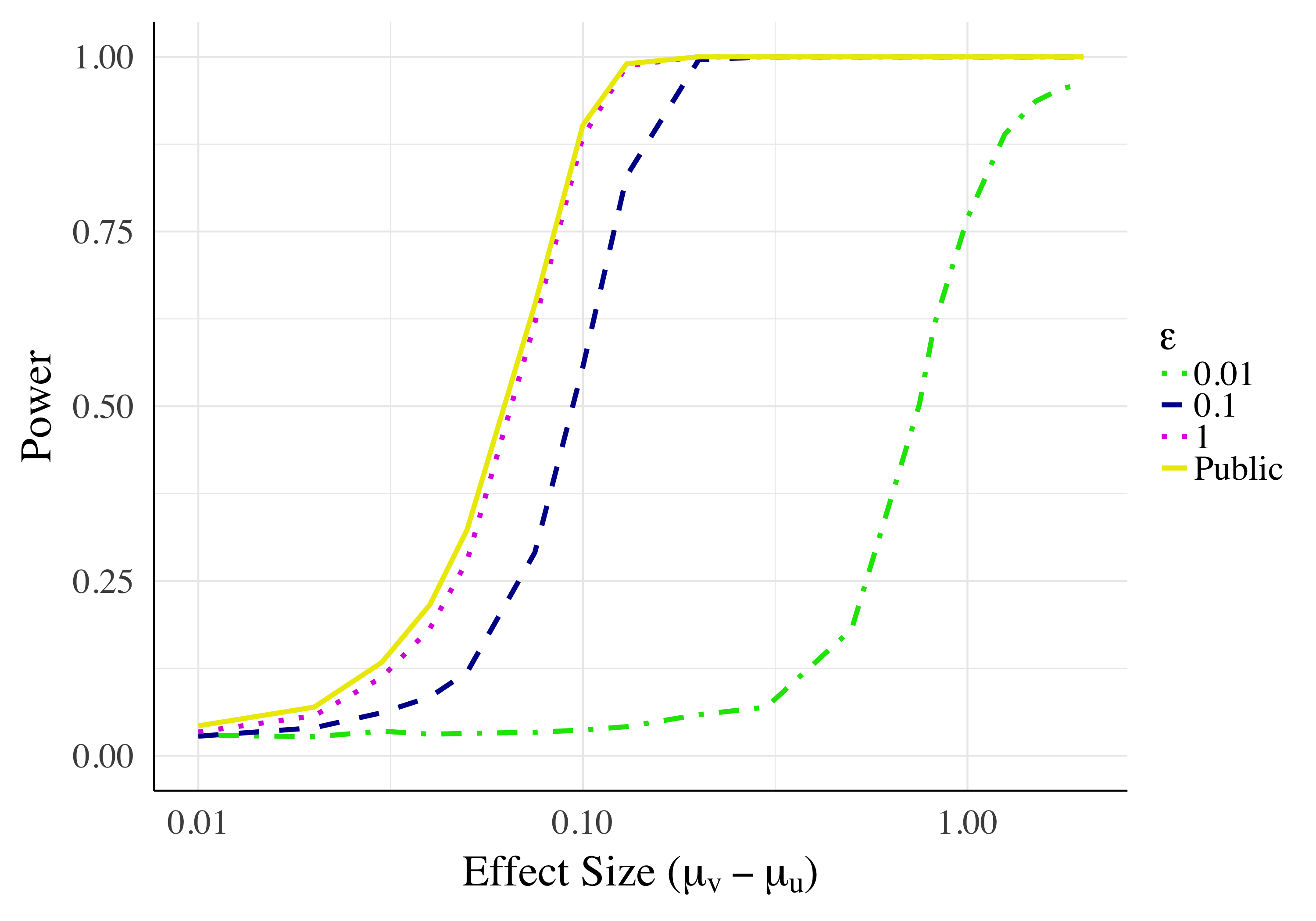}
    \caption{Power at various values of $\epsilon$ and effect size. ($\alpha = .05$, $n = 2500$)}
    \label{fig:varying_effect_size}
\end{figure}

We also examine the power of our test at varying effect sizes. As shown in Figure \ref{fig:varying_effect_size}, at large sample sizes $n$, in this case $2500$, there is essentially no difference between the minimum effect detectable in the private and public setting.  This is because the random variation in the sample overwhelms the relatively small random noise being added for privacy.  We also note, while not shown in the figure, that for small enough choices of $n$ and $\epsilon$ no effect, no matter how large, can be detected.  (Once all $d_i$ values are positive, increasing the effect size further has no effect.)

%% file: Comparison.tex
\section{Comparison to Previous Work}\label{comp_sec}

In 2016, Task and Clifton \cite{DPWilcoxon} introduced the first differentially private version of the Wilcoxon signed-rank test, from here on referred to as the TC test. Our work improves upon their test in two ways. We describe the two key differences below, and then compare the power of our test to theirs.  We also found a significant error in their work.\footnote{This error has been confirmed by Task and Clifton in personal correspondence.}  All comparisons are made to our implementation of the TC test with the relevant error corrected.

\paragraph{Computing critical values}  Task and Clifton compute an analytic upper bound on the critical value $t^*$.  For a given $n$ and $\epsilon$, the private test statistic $\widetilde{W}$ under $H_0$ is equal to a sum $W + \Lambda$, where $W$ is a random draw from a normal distribution (scaled according to $n$) and $\Lambda$ is a Laplace random variable (scaled according to $n$ and $\epsilon$).  In particular, say that $b$ is a value such that $\Pr[W>b] < \beta$ and $g$ is a value such that $\Pr[\Lambda > g] < \gamma$.  Then we can compute the following bound.  (The last line follows from the fact that the two events are independent.)
\begin{align*}
\Pr[\widetilde{W} > b + g] &< \Pr[W > b \text{ or } \Lambda > g] \\
&= \Pr[W > b] + \Pr[\Lambda > g]\\ 
&\ \ \ - \Pr[W > b \text{ and } \Lambda > g] \\
&= \beta + \gamma - \beta\gamma
\end{align*}
Task and Clifton always set $\gamma = .01$ and then vary  the choice of $\beta$ such that they have $\alpha = \beta + \gamma - \beta\gamma$ for whatever $\alpha$ is intended as the significance threshold.  (This is where Task and Clifton make an error.  This formula is correct, but they used an incorrect density function for the Laplace distribution and as a result calculated incorrect values of $g$.)

The bound described above is correct but is very loose.  We instead compute the critical value by simulation.  We use 10 million draws from the relevant distribution and compute the $1-\alpha$ quantile of their absolute values.  This gives drastically lower critical values.  Table \ref{tab:n_100_comp} contains examples of the critical values achieved by each method for several parameter choices. More values can be found in Appendix \ref{more_figs}.

\input{n_100_crit_val_comp_table}

\paragraph{Handling zero values} Our second key change from the TC test is that we handle rows with $d_i=0$ according to the Pratt variant of the Wilcoxon, rather than dropping them completely as is more traditional.  The reason the traditional method is so difficult in the private setting is that the reference distribution one must compare to depends on the number of rows that were dropped.  If $n_r$ is the number of non-zero rows (i.e., rows that weren't dropped), one is supposed to look up the critical value associated with $n_r$, rather than the original size $n$ of the database.  

Unfortunately, $n_r$ is a sensitive value and cannot be released privately.\footnote{A private estimate could be released, but one would have to devote a significant portion of the privacy budget for the hypothesis test to this estimate, greatly decreasing the accuracy/power of $\privW$.}  Task and Clifton show that it is \textit{acceptable} (in that it does not result in type I error greater than $\alpha$) to compare to a critical value for a value of $n_r$ that is lower than the actual value.  This allows them to give two options for how one might deal with the lack of knowledge about $n_r$.
\begin{description}
\item[High Utility] This version of the TC test simply assumes $n_r \geq .3n$ and uses the critical value that would be correct for $n_r = .3n$.  We stress that this algorithm is \textit{not} actually differentially private, though it could easily be captured by a sufficiently weakened definition that limited the universe of allowable databases.  Another problem is that for most realistic data, $n_r$ is much greater than $.3n$ and using this loose lower bound still results in a large loss of power.
\item[High Privacy] This version adds $k$ dummy values to the database with $d_i = \infty$ and $k$ with $d_i = -\infty$.\footnote{Task and Clifton do not discuss how to choose $k$, and in our experimental comparisons we set $k=15$, the same value they use.}  Then one can be certain of the bound $n_r \geq 2k$.  This is a guaranteed bound so this variant truly satisfies differential privacy.  On the other hand it is a very loose lower bound in most cases, leading to a large loss of power. 
\end{description}

\begin{figure} [!htb] 
    \centering
    \includegraphics[width=\linewidth]{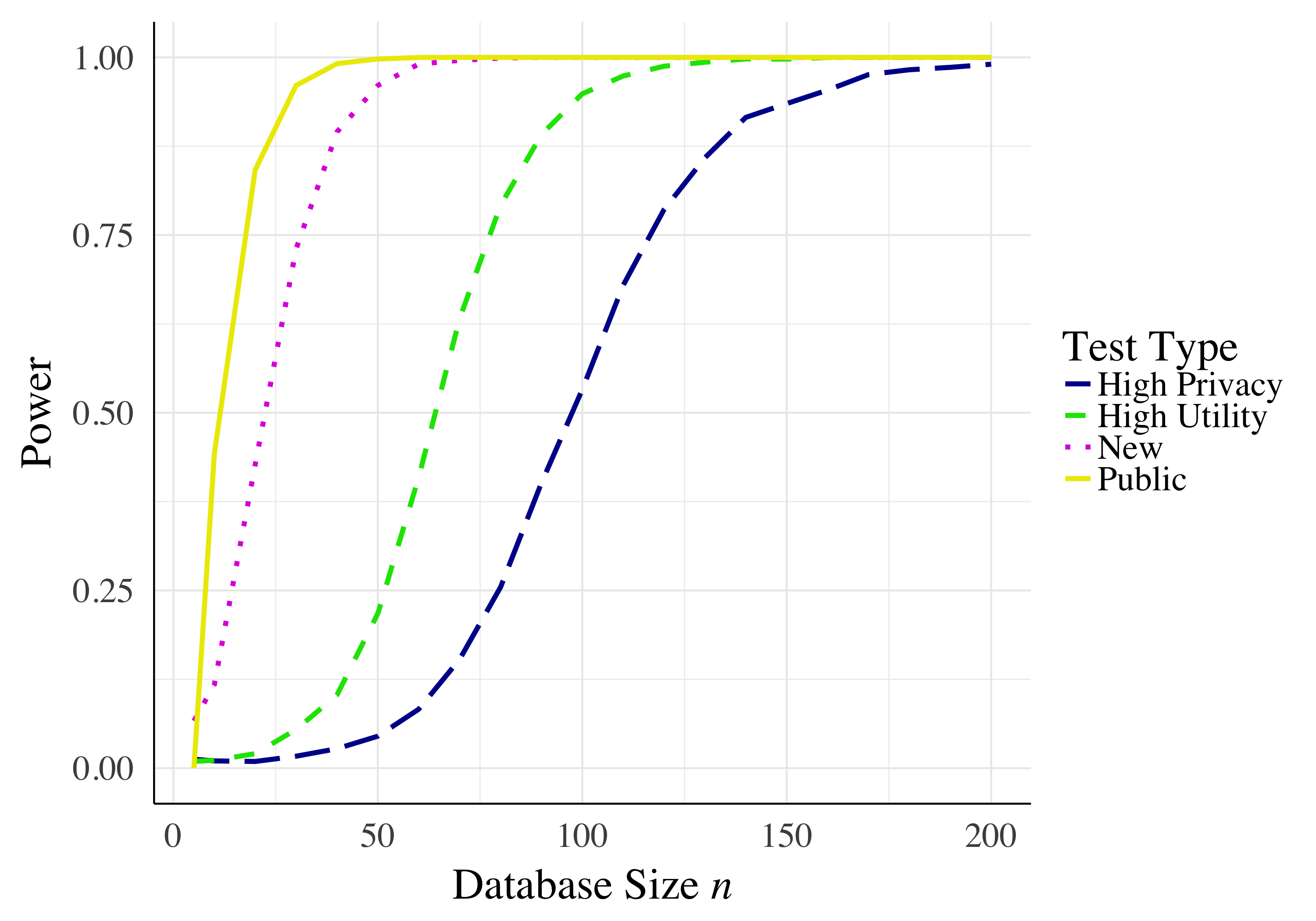}
    \caption{Power comparison of TC's \protect \cite{DPWilcoxon} algorithms, our new algorithm, and the public algorithm at various $n$. (Effect size: $\mu_u - \mu_v = 1 \sigma$, $\epsilon = 1$; $\alpha=.05$})
    \label{fig:comp_no_ties}
\end{figure}

\paragraph{Experimental results}  We compare our test to the TC test by measuring statistical power, just as we compared it to the public test in Section \ref{results}.  We begin by again measuring the power when detecting the difference between two normal distributions with means one standard deviation apart.  The results can be seen in Figure \ref{fig:comp_no_ties}.  If we look at the database size needed to achieve 80\% power, we find that the 32 data points we need, while more than the public test (14), are many fewer than the TC high utility variant (80) or the TC high privacy variant (122).  Appendix \ref{more_figs} includes a figure for $\epsilon=.01$ as well.  What we see is that, while all private tests require more data, our test (requiring $n\approx 236$) still requires about 40\%  as much data as the TC high utility variant (588).  The TC high privacy variant, however, scales much less well to low $\epsilon$ and requires roughly 2974 data points.

\begin{figure} [!htb] 
    \centering
    \includegraphics[width=\linewidth]{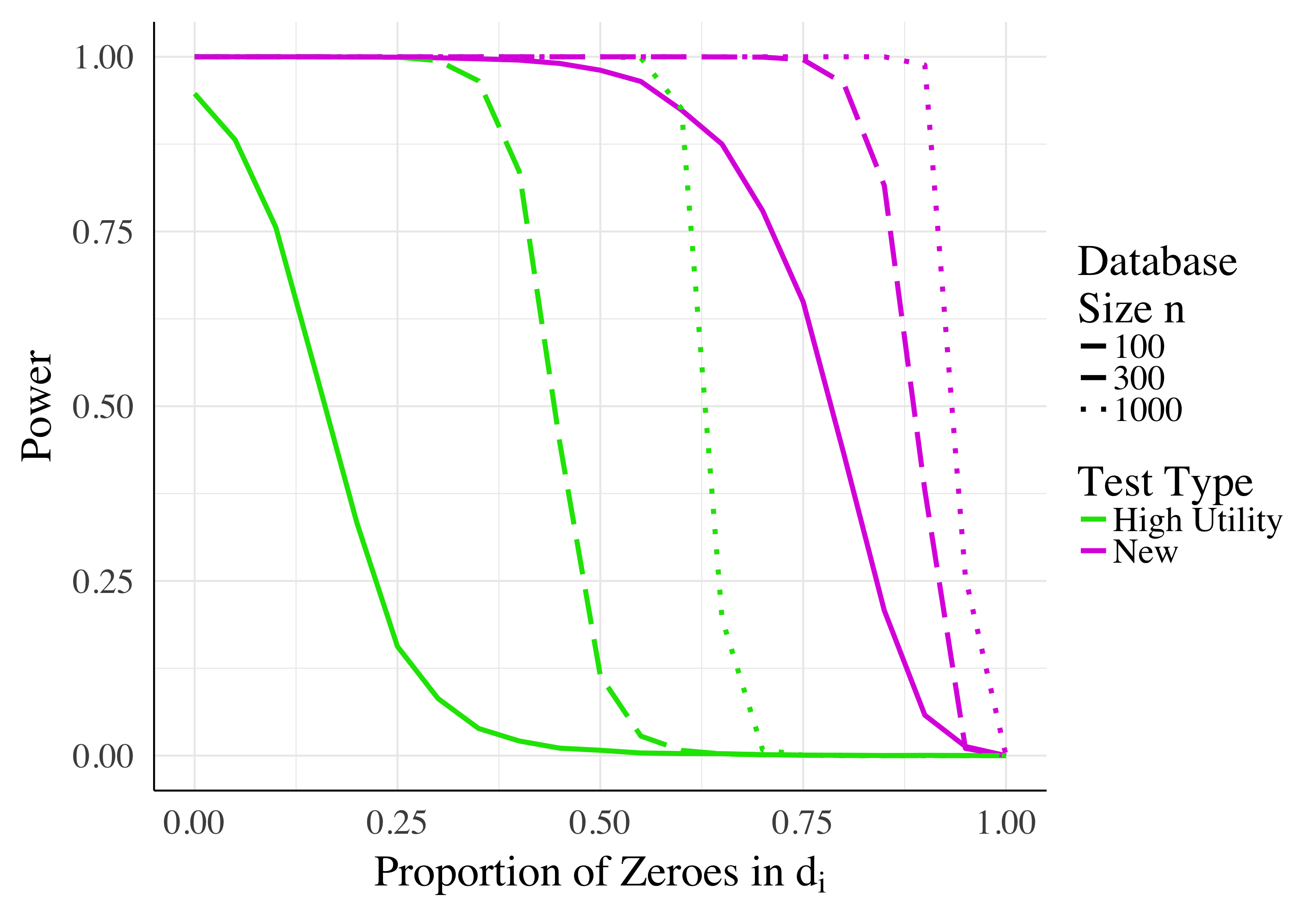}
    \caption{Power comparison of the TC \textit{High Utility} algorithm and our algorithm at various proportions of tied values and sample sizes $n$. (Effect size: $\mu_u - \mu_v = 1 \sigma$; $\epsilon = 1$; $\alpha = .05$)}
    \label{fig:ties-utility}
\end{figure}

\begin{figure} [!htb] 
    \centering
    \includegraphics[width=\linewidth]{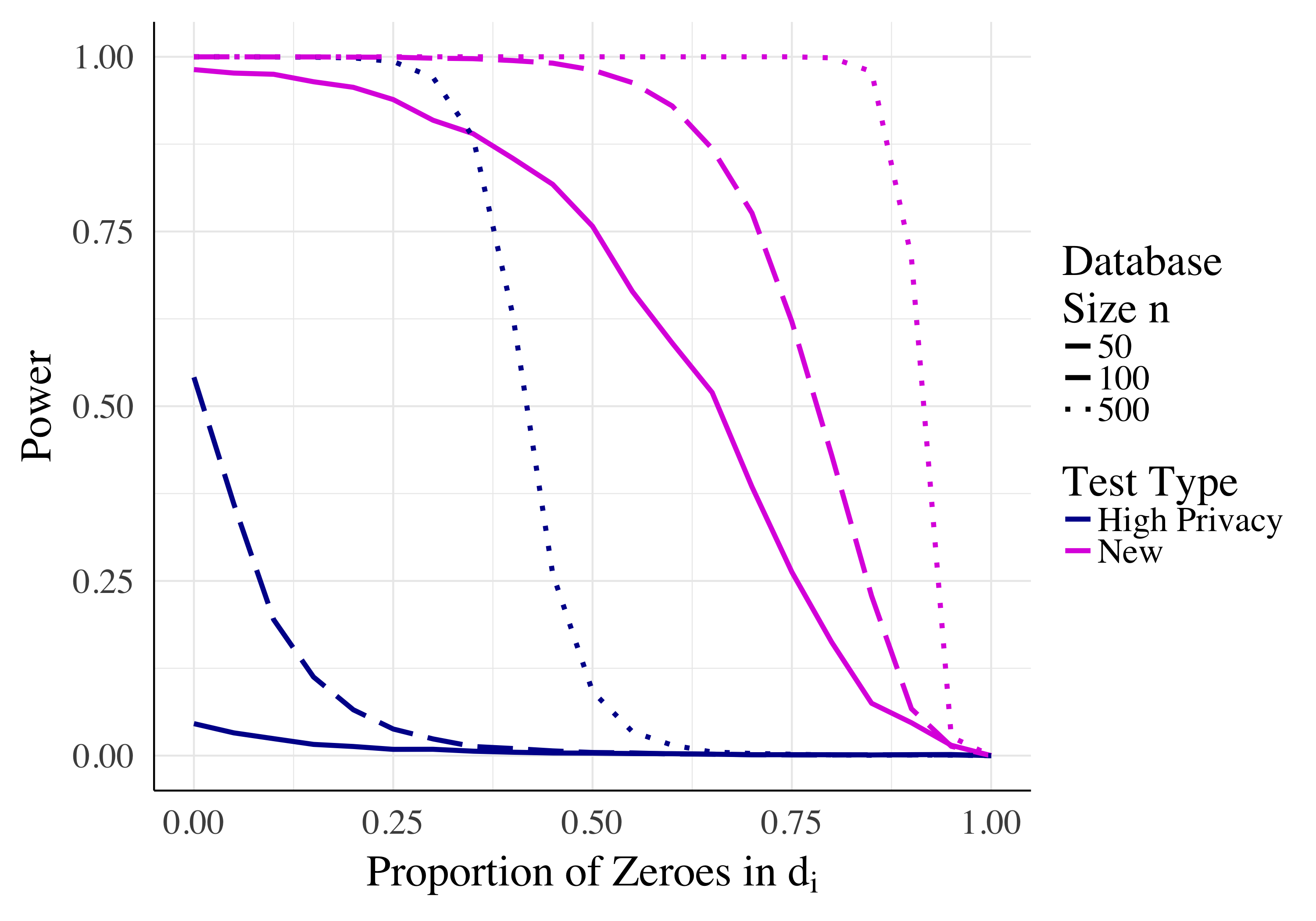}
    \caption{Power comparison of the TC \textit{High Privacy} algorithm ($k = 15$) and our algorithm at various proportions of tied values and sample sizes $n$. (Effect size: $\mu_u - \mu_v = 1 \sigma$; $\epsilon = 1$; $\alpha = .05$)}
    \label{fig:ties-privacy}
\end{figure}

The results in Figure \ref{fig:comp_no_ties} use a continuous distribution for the real data, so there are no data points with $d_i=0$.  Because one of the crucial differences between our algorithms is the method for handling these zero values, we also consider the effect when there are a large number of zeros.  In particular, Figures \ref{fig:ties-utility} and \ref{fig:ties-privacy} compare our algorithm to the high utility and high privacy variants of the TC test, respectively.  Here we first choose the number of rows with $d_i=0$ and then sample the remaining data points as before.  Of course, as the number of rows showing no difference increases, the power of all tests decreases, but we see that in all cases our test retains power longer.

Overall, we see that both in situations with no zero values and situations with many, our test achieves the rigorous privacy guarantees of the TC high privacy test while achieving greater utility than the TC high utility test.

\begin{figure} [!htb] 
    \centering
    \includegraphics[width=\linewidth]{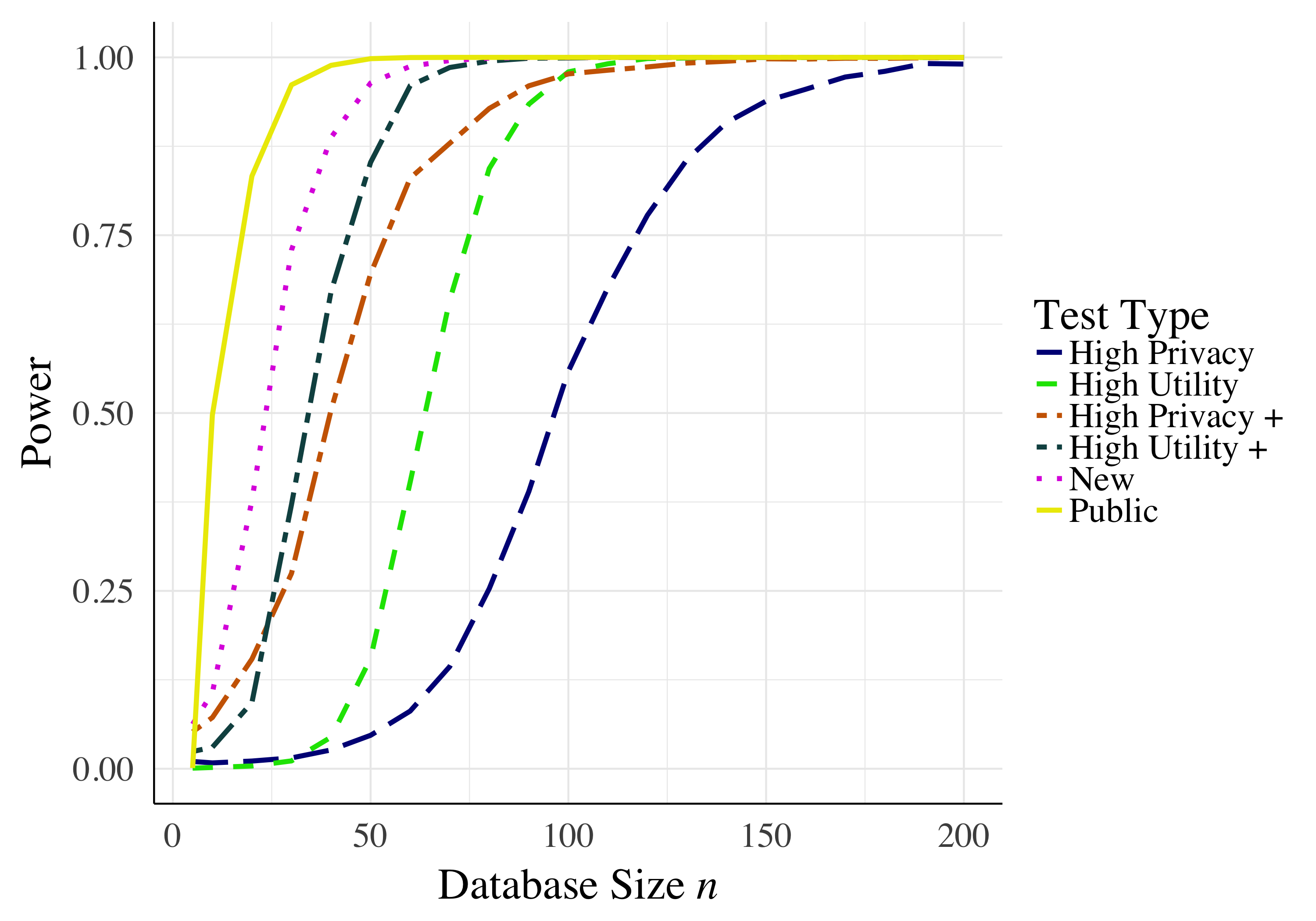}
    \caption{Power comparison of the TC algorithms, the TC algorithms with our critical values (denoted with a +), our new algorithm, and the public algorithm at various sample sizes $n$. (Effect size: $\mu_u - \mu_v = 1 \sigma$; $\epsilon=1$; $\alpha=.05$)}
    \label{fig:comp_no_ties_w_plus}
\end{figure}

\paragraph{Relative contribution of improvements}
Given that we make two meaningful changes to the TC test, one might naturally wonder whether both are truly useful or whether the vast majority of the improvement comes from one of the two changes.  To test this, we compare to an updated variant of the TC test where we calculate critical values exactly through simulation, as we do in our algorithm, but otherwise run the TC test unchanged. The result is presented in Figure \ref{fig:comp_no_ties_w_plus}. "High Privacy +" and "High Utility +", in Figure \ref{fig:comp_no_ties_w_plus}, refer to algorithms using our critical value computation and their methods of handling data as stated before.

We find the resulting algorithm to rest comfortably between the original TC test and our proposed test.  This means that both the change to the critical value calculation and the switch to the Pratt method of handling $d_i=0$ rows are important contributions to achieving the power of our test.

%% file: n_100_crit_val_comp_table.tex
% latex table generated in R 3.4.1 by xtable 1.8-2 package
% Wed Jul 11 13:40:40 2018
\begin{table}[ht]
\centering
\caption{Critical Value Comparison for $n=100$}\label{tab:n_100_comp}
\begin{tabular}{llrrr}
  \hline
$\epsilon$ & $\alpha$ & Public & New & TC \\ 
  \hline
1 & 0.1 & 1.282 & 1.417 & 2.680 \\ 
   & 0.05 & 1.645 & 1.826 & 3.091 \\ 
   & 0.025 & 1.960 & 2.186 & 3.511 \\ 
   \hline
0.1 & 0.1 & 1.282 & 5.684 & 14.786 \\ 
   & 0.05 & 1.645 & 8.063 & 15.197 \\ 
   & 0.025 & 1.960 & 10.438 & 15.617 \\ 
   \hline
0.01 & 0.1 & 1.282 & 55.350 & 135.843 \\ 
   & 0.05 & 1.645 & 79.233 & 136.254 \\ 
   & 0.025 & 1.960 & 103.116 & 136.674 \\ 
   \hline
\end{tabular}
\\[10pt]
\caption*{Critical values for \textit{n} = 100 and several values of $\epsilon$ and $\alpha$.  To allow easy comparison, these values are for a normalized $W$ statistic, i.e., $W$ has been divided by the relevant constant so that it is (before the addition of Laplacian noise) distributed according to a standard normal. See Appendix \ref{more_figs} for the equivalent table at n = 1000.}
\end{table}

%% file: Application.tex
\section{Application to Real-World Data} \label{realdata}

We now demonstrate the use of our algorithm on real-world data.  We use a database of New York City tax ride information released in 2014.  The database contains information on every Yellow Taxi ride in New York City in 2013, and its release resulted in high-profile de-anonymization attacks \cite{NYCData}.  Our main finding is that (at least for some natural analyses that we attempt) a differentially private query interface would have been sufficient, and the release of the data set was unnecessary.  We also again compare our statistical power to that of the TC tests and find it superior.\footnote{This data set is the same one Task and Clifton originally use for evaluating their test.}

This dataset contains several variables of interest for every Yellow Taxi ride in New York City in 2013, of which the following will be most useful:

\begin{itemize}
  \item Hack License: a unique identifier for every taxi driver in the city
  \item Number of Passengers: how many people rode together in the taxi
  \item Trip Time: the duration of the ride in seconds
  \item Trip Distance: the total distance traveled during the ride
\end{itemize}

We subsetted this dataset, initially, to include rides occuring on January 1st and 2nd of 2013. Then, we only kept rides given by a driver who drove on both the 1st and the 2nd.  In one data set, $u_i$ and $v_i$ were the average trip time on the 1st and 2nd, respectively.  The Wilcoxon test can then be used to test whether trip time varied between the two days. We calculated two other data sets similarly for trip distance and number of passengers.

\begin{figure} [!htb] 
    \centering
    \includegraphics[width=\linewidth]{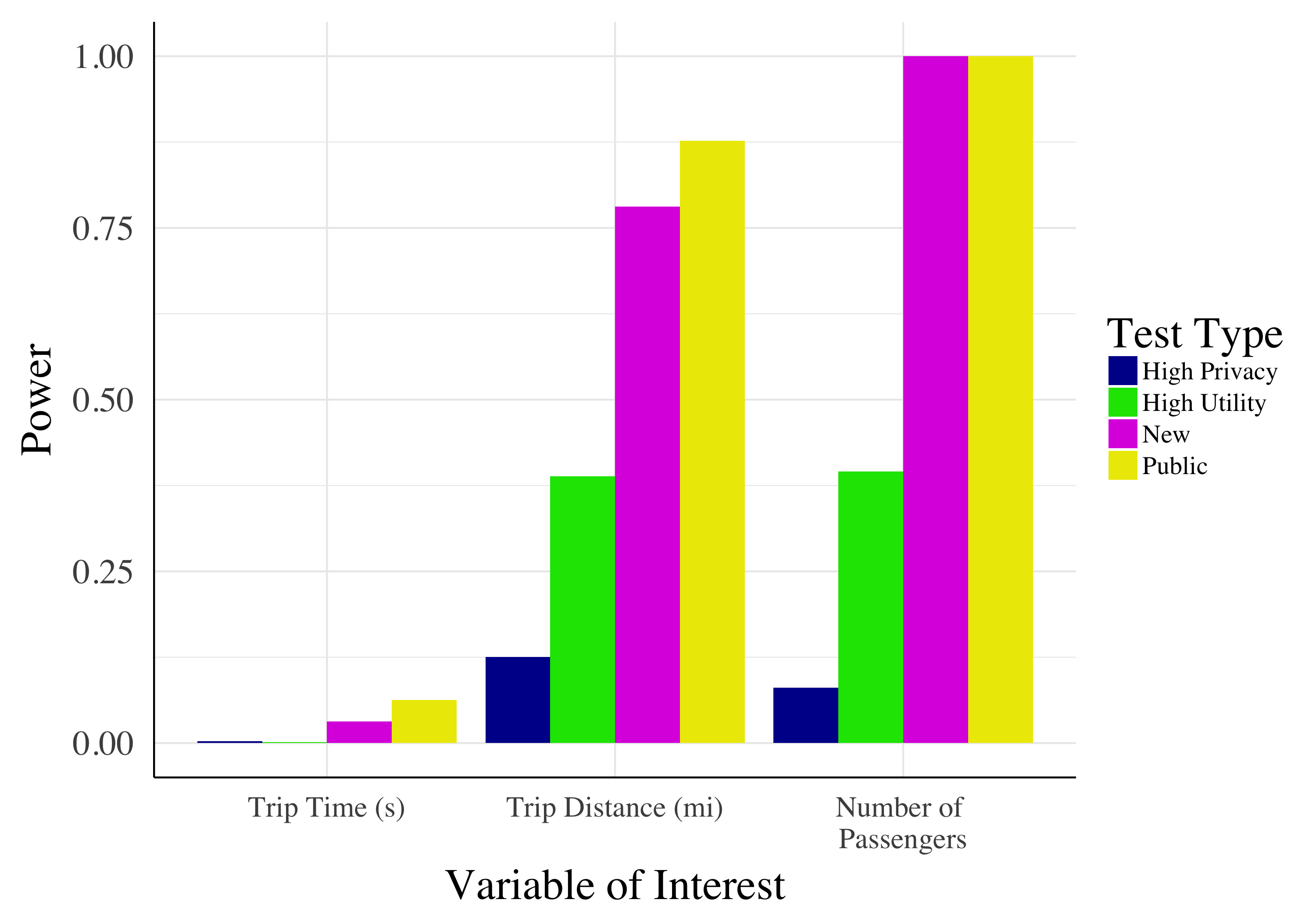}
    \caption{Power of the TC tests, our test, and the public test at sample sizes of 400, sampled with replacement from the NYC taxi data. ($\epsilon = 1$; $\alpha = .05$)}
    \label{fig:App}
\end{figure}

The resulting data sets have 17,066 entries.  We sampled (with replacement) $10^5$ different smaller datasets of size $n=400$.  Finally, we ran each algorithm on each of these resulting data sets and report the proportion in which a significant result (at $\alpha = .05$) was found. The results are summarized in Figure \ref{fig:App}. 

The power of our test follows the public test closely, and we see that for a data set as small as 400, we can achieve results with a private query that are almost as useful as a full public release of the data.  The TC algorithms do not achieve this goal, though the high utility variant is still meaningfully useful.  Of course, there will be choices of $n$ and $\epsilon$ for which the gap between our power and the power of the public test is quite large, but our point here is to argue that for even reasonably small data sets a private query interface can be sufficient for many important tasks.

%% file: Conclusion.tex
\section{Conclusion and Open Questions}

We have shown a new private variant of the Wilcoxon hypothesis test complete with an accurate method for computing p-values.  It contains two significant technical innovations over prior work.  We show that this test has much higher power than the previously available tests, and (at least for $\epsilon=1$) its power is actually not very far from that of the public test.  We also show that it is a realistic algorithm to run on at least some real-world data sets at reasonable database sizes for many applications.  We think this is a major step forward, but several questions remain open.

The most obvious goal is simply to continue improving the power of the test.  The gap between our test and the public test is still significant when $\epsilon$ is meaningfully less than 1.  

It would also be useful to compare this test to other tests in the statistics literature.  For example, if one knows that the underlying data is drawn from a normal distribution, classical statistics would use a paired-sample t-test, which is known to have higher power in this case than the Wilcoxon.  We suspect that the Wilcoxon test holds up better under the constraints of differential privacy and that it might actually be better in the private setting to use the private Wilcoxon \textit{even when} the data is known to be normally distributed.  Unfortunately, to our knowledge no one has yet developed a private analogue of the t-test, so we are unable to make the comparison.\footnote{This material is based upon work supported by the National Science Foundation under Grant No. SaTC-1817245 and the Richter Funds.}

%% file: Appendix.tex
\section{Uniformity of p-values}\label{uniformp}

The uniformity of p-values for a hypothesis test can be assessd by simulating many
p-values and evaluating how well they follow the uniform distribution on the unit interval. A common tool for this purpose is the quantile-quantile (or Q-Q) plot, which plots the quantiles of the theoretical distribution (the uniform in this case) against the empirical quantiles of the p-values. A sample that perfectly follows the theoretical distribution will coincide at all quantiles and be represented on the Q-Q plot as the identity line.

\begin{figure} [!htb] 
    \centering
    \includegraphics[width=\linewidth]{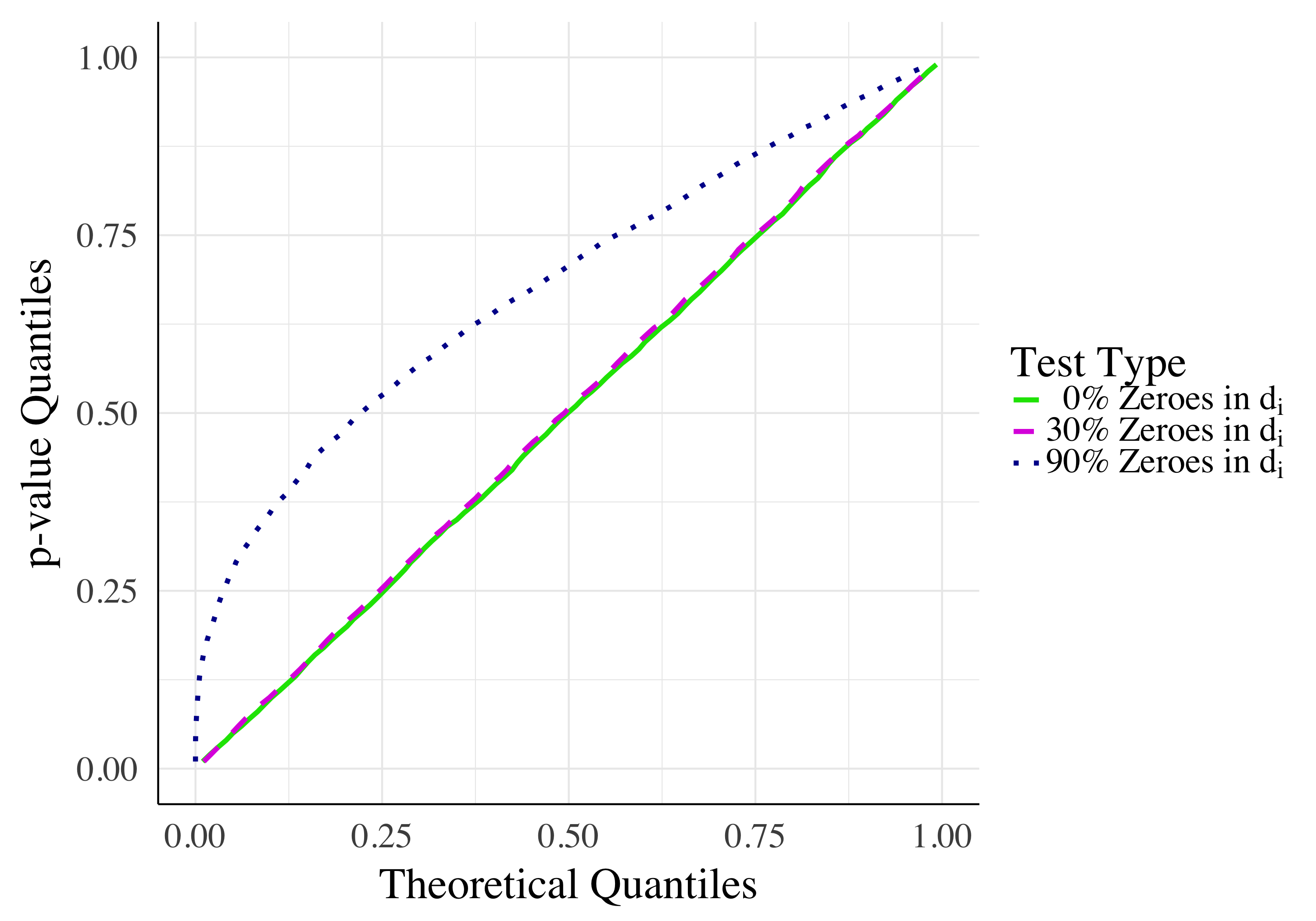}
    \caption{A quantile-quantile plot comparing the distribution of simulated p-values to the uniform distribution ($\epsilon = 1$, $n = 500$).}
    \label{fig:qqplot}
\end{figure}

Figure \ref{fig:qqplot} shows a Q-Q plot of three sets of p-values, all generated under $H_0$, with $\epsilon = 1$, $n = 500$. When there are no ties in the original data (0\% of $d_i = 0$), the Q-Q plot line is indistinguishable from the identity line, indicating that the test is properly calibrated. Encouragingly, introducing a substantial number of ties into the data (30\% of $d_i = 0$) has no noticible effect.

In order to induce non-uniformity in the p-values, one needs an extremely high proportion of rows with $d_i=0$.  The curve with 90\% zero values is shown as an illustration.  When the proportion of zeros is very high, the variance of the p-values will be narrower than the reference distribution, resulting in a lower critical value.  Since the value we are using is higher, our test is overly conservative,\footnote{One could try to estimate the number of zeros to be less conservative, but that would require allocating some of the privacy budget towards that estimate, which is not worth it in most circumstances.} but this is acceptable as type I error is still below $\alpha$.

\clearpage
\section{Additional Figures}\label{more_figs}

\subsection{Power Comparison}

\begin{figure} [!htb] 
    \centering
    \includegraphics[width=\linewidth]{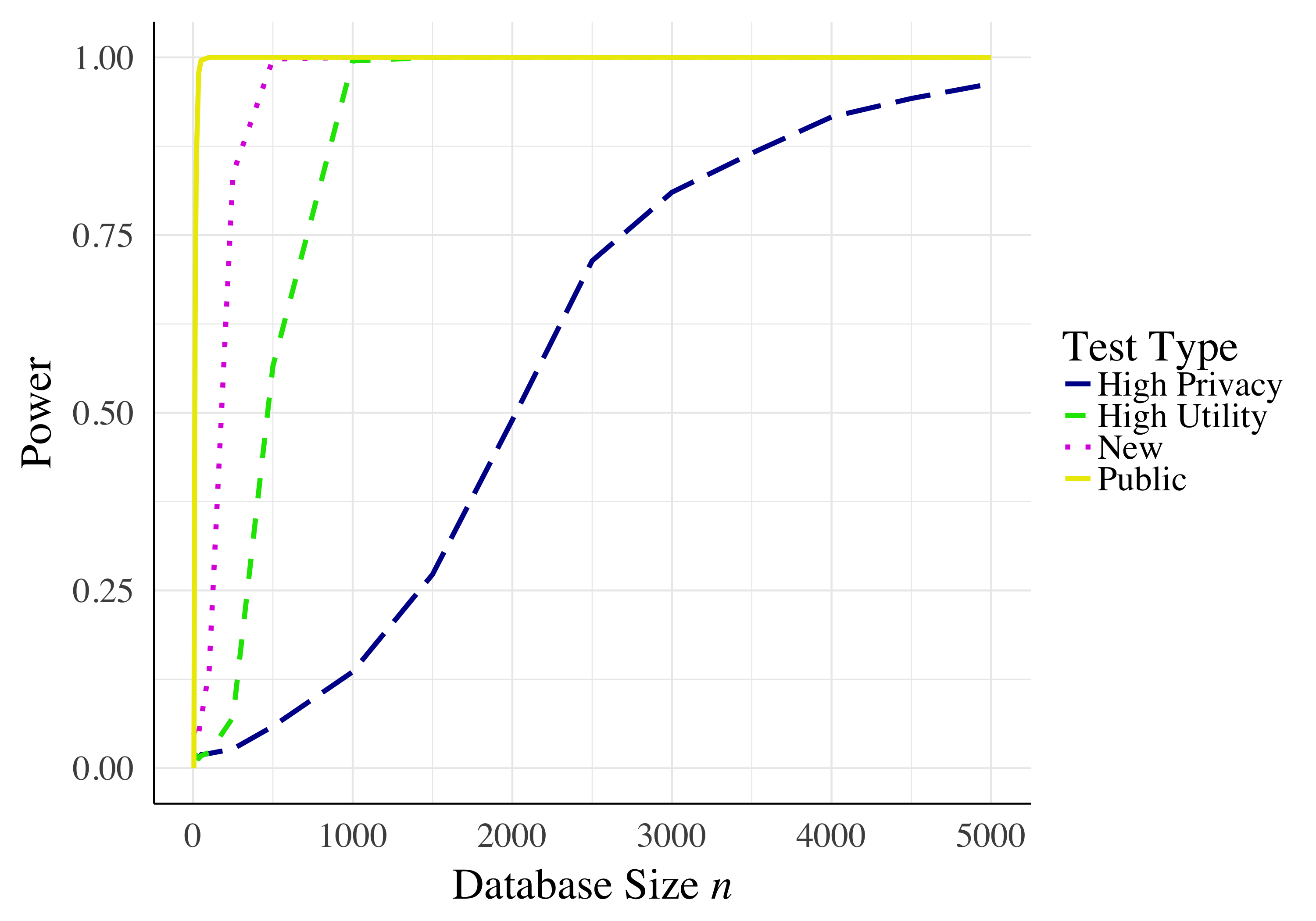}
    \caption{Power comparison of Task and Clifton's algorithms, our new algorithm, and the public algorithm at various database sizes $n$. (Effect size: $\mu_u - \mu_v = 1 \sigma$; $\epsilon=.1$; $\alpha=.05$)}
    \label{fig:comp_low_eps}
\end{figure}

\subsection{Critical Value Tables}

\input{n_1000_crit_val_comp_table.tex}

\input{crit_val_eps_1p0.tex}

\input{crit_val_eps_0p1.tex}

\input{crit_val_eps_p01.tex}

%% file: n_1000_crit_val_comp_table.tex
% latex table generated in R 3.4.1 by xtable 1.8-2 package
% Wed Jul 11 13:40:40 2018
\begin{table}[ht]
\centering
\caption{Critical Value Comparison for $n=1000$}\label{tab:n_1000_comp}
\begin{tabular}{llrrr}
  \hline
$\epsilon$ & $\alpha$ & Public & New & TC \\ 
  \hline
1 & 0.1 & 1.282 & 1.296 & 1.763 \\ 
   & 0.05 & 1.645 & 1.665 & 2.174 \\ 
   & 0.025 & 1.960 & 1.984 & 2.594 \\ 
   \hline
0.1 & 0.1 & 1.282 & 2.203 & 5.617 \\ 
   & 0.05 & 1.645 & 2.975 & 6.028 \\ 
   & 0.025 & 1.960 & 3.740 & 6.448 \\ 
   \hline
0.01 & 0.1 & 1.282 & 17.681 & 44.157 \\ 
   & 0.05 & 1.645 & 25.234 & 44.568 \\ 
   & 0.025 & 1.960 & 32.844 & 44.988 \\ 
   \hline
\end{tabular}
\\[10pt]
\caption*{Critical values for \textit{n} = 1000 and several values of $\epsilon$ and $\alpha$.  To allow easy comparison, these values are for a normalized $W$ statistic, i.e., $W$ has been divided by the relevant constant so that it is (before the addition of Laplacian noise) distributed according to a standard normal.}
\end{table}

%% file: crit_val_eps_1p0.tex
% latex table generated in R 3.4.1 by xtable 1.8-2 package
% Tue Jul 10 15:43:38 2018
\begin{table}[ht]
\centering
\caption{New Critical Value Table for $\epsilon$ = $1.0$}
\label{tab:crit_val_eps_1.0}
\begin{tabular}{rrrrr}
  \hline
$n$ & 0.05 & 0.025 & 0.01 & 0.005 \\ 
  \hline
10 & 70 & 83 & 102 & 116 \\ 
  20 & 155 & 183 & 220 & 248 \\ 
  30 & 256 & 299 & 355 & 397 \\ 
  40 & 369 & 429 & 506 & 562 \\ 
  50 & 494 & 572 & 670 & 742 \\ 
  75 & 854 & 984 & 1143 & 1257 \\ 
  100 & 1271 & 1460 & 1690 & 1853 \\ 
  200 & 3402 & 3895 & 4486 & 4900 \\ 
  300 & 6127 & 7012 & 8069 & 8798 \\ 
  400 & 9335 & 10679 & 12276 & 13382 \\ 
  500 & 12978 & 14845 & 17061 & 18592 \\ 
  1000 & 36235 & 41443 & 47637 & 51906 \\ 
   \hline
\end{tabular}
\\[10pt]
\caption*{Critical values at several sample sizes $n$ and two-sided significance levels $\alpha$. To calcuate these values, we run 10 million simulations for each parameter combination and compute the $1$ $-$ $\alpha$th percentile of the absolute value of the distribution.}
\end{table}

%% file: crit_val_eps_0p1.tex
% latex table generated in R 3.4.1 by xtable 1.8-2 package
% Tue Jul 10 15:43:39 2018
\begin{table}[ht]
\centering
\caption{New Critical Value Table for $\epsilon$ = $0.1$}
\label{tab:crit_val_eps_0.1}
\begin{tabular}{rrrrr}
  \hline
$n$ & 0.05 & 0.025 & 0.01 & 0.005 \\ 
  \hline
10 & 600 & 739 & 922 & 1061 \\ 
  20 & 1202 & 1479 & 1846 & 2123 \\ 
  30 & 1806 & 2220 & 2770 & 3185 \\ 
  40 & 2413 & 2968 & 3704 & 4261 \\ 
  50 & 3018 & 3713 & 4628 & 5324 \\ 
  75 & 4541 & 5577 & 6954 & 7989 \\ 
  100 & 6073 & 7461 & 9294 & 10677 \\ 
  200 & 12328 & 15098 & 18767 & 21531 \\ 
  300 & 18733 & 22892 & 28391 & 32519 \\ 
  400 & 25296 & 30837 & 38193 & 43736 \\ 
  500 & 32054 & 38979 & 48128 & 55083 \\ 
  1000 & 68258 & 82120 & 100408 & 114230 \\ 
   \hline
\end{tabular}
\\[10pt]
\caption*{Critical values at several sample sizes $n$ and two-sided significance levels $\alpha$. To calcuate these values, we run 10 million simulations for each parameter combination and compute the $1$ $-$ $\alpha$th percentile of the absolute value of the distribution.}
\end{table}

%% file: crit_val_eps_p01.tex
% latex table generated in R 3.4.1 by xtable 1.8-2 package
% Tue Jul 10 15:43:40 2018
\begin{table}[ht]
\centering
\caption{New Critical Value Table for $\epsilon$ = $0.01$}
\label{tab:crit_val_eps_.01}
\begin{tabular}{rrrrr}
  \hline
$n$ & 0.05 & 0.025 & 0.01 & 0.005 \\ 
  \hline
10 & 5992 & 7377 & 9209 & 10596 \\ 
  20 & 11971 & 14742 & 18416 & 21196 \\ 
  30 & 17976 & 22137 & 27644 & 31774 \\ 
  40 & 23974 & 29516 & 36877 & 42425 \\ 
  50 & 29964 & 36905 & 46081 & 53034 \\ 
  75 & 44933 & 55371 & 69105 & 79513 \\ 
  100 & 59921 & 73792 & 92066 & 106005 \\ 
  200 & 119902 & 147619 & 184222 & 212010 \\ 
  300 & 179942 & 221477 & 276678 & 317895 \\ 
  400 & 239695 & 295106 & 368374 & 423528 \\ 
  500 & 299627 & 368763 & 460256 & 529522 \\ 
  1000 & 600096 & 738071 & 921529 & 1061150 \\ 
   \hline
\end{tabular}
\\[10pt]
\caption*{Critical values at several sample sizes $n$ and two-sided significance levels $\alpha$. To calcuate these values, we run 10 million simulations for each parameter combination and compute the $1$ $-$ $\alpha$th percentile of the absolute value of the distribution.}
\end{table}